# Embedding Recurrent Layers with Dual-Path Strategy in a Variant of Convolutional Network for Speaker-Independent Speech Separation

*Xue Yang, Changchun Bao*[*]

Speech and Audio Signal Processing Laboratory, Faculty of Information Technology,
Beijing University of Technology, Beijing, China
yangx11@emails.bjut.edu.cn, baochch@bjut.edu.cn

## Abstract

Speaker-independent speech separation has achieved remarkable performance in recent years with the development of deep neural network (DNN). Various network architectures, from traditional convolutional neural network (CNN) and recurrent neural network (RNN) to advanced transformer, have been designed sophistically to improve separation performance. However, the state-of-the-art models usually suffer from several flaws related to the computation, such as large model size, huge memory consumption and computational complexity. To find the balance between the performance and computational efficiency and to further explore the modeling ability of traditional network structure, we combine RNN and a newly proposed variant of convolutional network to cope with speech separation problem. By embedding two RNNs into basic block of this variant with the help of dual-path strategy, the proposed network can effectively learn the local information and global dependency. Besides, a four-staged structure enables the separation procedure to be performed gradually at finer and finer scales as the feature dimension increases. The experimental results on various datasets have proven the effectiveness of the proposed method and shown that a trade-off between the separation performance and computational efficiency is well achieved.

## 1. Introduction

As an important part of front-end speech processing, speech separation has become one of the hot topics in recent years and has witnessed major advances in performance. While human beings have the impressive ability to separate clean speech in complex acoustic scenarios (e.g.: with the presence of interfering speakers and/or background noises), it seems really difficult for machine system to tackle this kind of task, which is widely known as the cocktail party problem [1]. Thus, speech separation aims to crack the problem by extracting individual speeches from the mixed signal [2].

After entering the era of deep learning, several algorithms and deep structures have been proposed to effectively realize speech separation for speaker-dependent or target-dependent situation [3-4]. However, for speaker-independent case, the primary permutation problem has remained unsolved until the advent of deep clustering (DPCL) [5] and permutation invariant training (PIT) [6]. Since then, advanced deep learning techniques have been applied to speaker-independent speech separation and the state-of-the-art performance has been improved dramatically as a result.

Typically, speech separation algorithms could be divided into two categories, i.e., frequency-domain methods and time-domain methods. The former usually copes with time-frequency (T-F) representation of the mixed signal obtained through the short-time Fourier transform (STFT). In order to approximate the clean T-F representation of each individual, the separation network either performs direct mapping that outputs T-F representation directly or performs mask estimation that will be multiplied by the representation of mixture to recover the T-F representation of each individual speaker. The estimated waveform is synthesized finally with the aid of inverse short-time Fourier transform (iSTFT) [7-9]. A major drawback of frequency-domain methods is that the mixed signal is actually transformed into a complex domain by STFT and the majority of separation algorithms only process the magnitude spectrogram, since no clear structure exists in the phase spectrogram [10]. Although the phase-sensitive mask (PSM) [11] and complex ideal ratio mask (cIRM) [10] were proposed to deal with the phase spectrogram, the reconstruction still could not perform exactly.

To overcome this difficulty, various time-domain methods were proposed and achieved excellent performance, in which several algorithms took advantage of the expressive power of deep neural networks and learned a regression function directly from the waveform of the mixture to the waveform of the underlying individual speaker [12-13]. Whereas some algorithms replaced STFT with the fixed or parameterized or learned transform and the mapped signal was further processed in a latent space with separation network [14-16], which is mainly used to estimate a mask for individual signal. Temporal convolutional neural network was applied in [17] so that the separation performance first surpassed ideal T-F magnitude masking. It is worth mentioning that the frame length was reduced to 2ms to obtain a learned transform more suitable for separation in this work. Because of the relatively small frame length compared to that used in frequency-domain methods, the ability of modeling long sequences effectively has become a primary problem and a variety of methods have boosted the state-of-the-art performance. Multi-resolution features and multi-scale fusion were adopted in [18-20] to effectively exploit information in various scales. In addition, the dual-path strategy was employed to process local information and global dependency iteratively in [21-23] and the separation network has been evolved from the traditional recurrent neural network (RNN) to the advanced transformer, which is notable for its attention mechanism to model long-term dependency. The frame length in [21-22] was further reduced to 2 samples at 8 kHz sampling rate for achieving a better performance. However, such a small frame length may result in an extremely long sequence. When RNN or modified transformer is utilized to model such a long sequence, the memory consumption and time-consuming are still a problem although the dual-path strategy was applied.

In this paper, inspired by a newly proposed variant of convolutional network structure, which is named as ConvNeXts and mainly applied in computer vision domain [24], we developed a novel recurrent convolutional network for speaker-independent speech separation. Since dual-path strategy [21] is also adopted to mitigate the problem introduced by long sequence modeling, we denote the proposed method as DP-RCNet. We find that the proposed network could achieve a trade-off between the separation performance and computational efficiency. In addition, we also evaluate our proposed method with more diverse datasets and the effectiveness is proved.

The rest of this paper is organized as follows. The general time-domain separation framework and detailed description of our proposed network structure are given in Section 2. Then we present the experimental setup in Section 3. Subsequently, the experimental results are showed and discussed in Section 4. Finally, we come to the conclusions in Section 5.

## 2. Architecture description

### 2.1. General framework

In anechoic situation, the mixed signal $y \in R^{T \times 1}$ can be expressed as the sum of $S$ individual signals $x_s \in R^{T \times 1}$ and additive noise signal $n \in R^{T \times 1}$, i.e.:

$$y = \sum_{s=1}^{S} x_s + n \quad (1)$$

where $T$ denotes the number of speech samples and $S$ denotes the total number of the speakers. Speech separation aims to estimate each individual signal $x_s$ from the mixture. The general framework of time-domain method [17] is depicted in Figure 1.

The general framework consists of an encoder, a separator and a decoder. The encoder splits the mixed signal $y$ into overlapping frames and maps the mixed signal of each frame $y^k \in R^{L \times 1}$ into a feature vector $f^k \in R^{N \times 1}$ in a latent space:

$$f^k = W_E y^k \quad (2)$$

where $k$ is the frame index, $L$ is the frame length, $N$ is the number of the features and $W_E \in R^{N \times L}$ corresponds to the weighting matrix of encoder. Usually, the encoder includes an activation function, e.g.: sigmoid, rectified linear unit (ReLU), etc. In this paper, no extra activation function is used in the encoder as in [17].

The separator takes the feature vectors as input and estimates a mask $M_s^k \in R^{N \times 1}$ for the $s^{th}$ speaker in the $k^{th}$ frame. The mask is multiplied by the feature vector $f^k$ of the mixed signal to estimate $\hat{f}_s^k$ of the $s^{th}$ speaker:

$$\hat{f}_s^k = M_s^k \odot f^k \quad (3)$$

where $\odot$ is the element-wise multiplication.

The decoder transforms the estimated feature vector $\hat{f}_s^k \in R^{N \times 1}$ back into time-domain signal $\hat{x}_s^k \in R^{L \times 1}$ for the $k^{th}$ frame, i.e.,

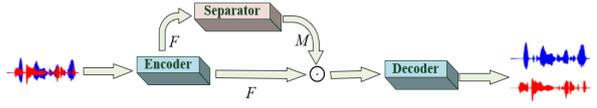

Figure 1: *General framework of speech separation.*

$$\hat{x}_s^k = W_D \hat{f}_s^k \quad (4)$$

where $W_D \in R^{L \times N}$ is the weighting matrix of decoder. Subsequently, the estimated signals by (4) in each frame are used to reconstruct speech signals $\hat{x}_s, s = 1,2,\ldots,S$ using the overlap-add operation. In practice, the encoder and decoder are realized by the 1-D convolution and 1-D transposed convolution, respectively. In this general framework, the separator plays an important role for controlling computational complexity and memory consumption. Current network structures such as excellent transformer have not gained a good trade-off between separation performance and computational efficiency. Therefore, we propose a novel separator which is detailed below.

### 2.2. The proposed separator

As shown in Figure 2, the proposed separator is mainly composed of three modules, i.e.: the segmentation module, the backbone module that consists of a series of the DP-RCNet blocks and the overlap-add module.

#### 2.2.1. Segmentation module

For a mixed signal, the encoder output is denoted as $F \in R^{N \times K}$, where $K$ is the total number of frames. This mapping sequence is fed into a normalization layer and the feature dimension is reduced to $B$ from $N$ through a 1-D convolutional layer with kernel size 1. The segmentation module is used to split the dimension-reduced sequence into the overlapped chunks with chunk size $I$ and hop size $H$. The resulting chunks are concatenated to form a stacked feature tensor $U \in R^{B \times I \times J}$, where the 1st dimension corresponds to feature dimension, the 2nd one corresponds to intra-chunk dimension and the 3rd one corresponds to inter-chunk dimension.

#### 2.2.2. Backbone module

The novel network structure ConvNeXts was intended to compete with transformers and has shown remarkable performance in various vision tasks [24]. The basic block in this novel structure is the ConvNeXt block, which is shown in Figure 3(a). Modified from the well-known ResNet block [25], the ConvNeXt block consists of three 2-D convolutional layers, one normalization layer and one activation function. The first convolution is depth-wise separable convolution [26] with large kernel size for increasing the receptive field. The second and third convolutions have opposite functionalities, which means that the second one enlarges the feature dimension by a factor of 4 and the third one brings it back by a factor of 0.25. Moreover, layer normalization and Gaussian error linear unit (GELU) [27] are applied.

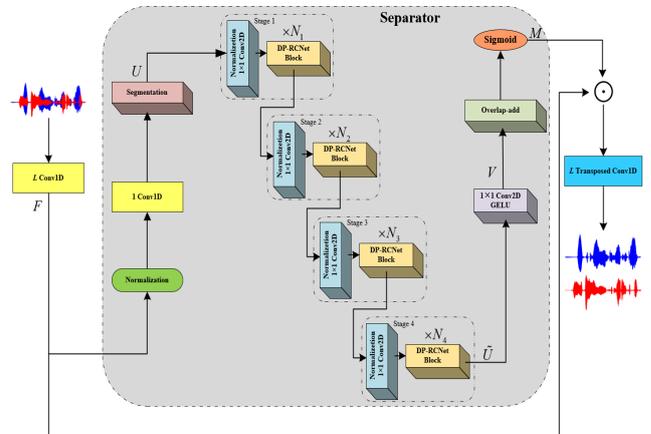

Figure 2: *The proposed DP-RCNet for speech separation.*

The original ConvNeXt block mainly deals with image related tasks. To better deal with sequential data, we propose to replace the first convolutional layer in ConvNeXt block with the RNNs. Since long sequence modeling is rather difficult, we adopt dual-path strategy [21] here to cascade two modified sub-blocks, one for modeling intra-chunk sequence and other one for modeling inter-chunk sequence. The resulting block is named as DP-RCNet block and is depicted in Figure 3(b).

As shown in Figure 2, the backbone module includes four stages of the DP-RCNet blocks. In each stage, the feature dimension in the successive DP-RCNet blocks keeps the same and is denoted as $D_r$, $r = 1, 2, 3, 4$, respectively. At the beginning of each stage, a $1 \times 1$ 2-D convolutional layer is used to vary the feature dimension and the feature dimension is doubled with the increase of stage, i.e.: $D_{r+1} = 2D_r$. The additional layer normalization before the convolutional layer in each stage is used to help stabilize the training as indicated in [24]. With this four-staged structure, the separation procedure could be performed gradually at finer and finer scales.

For each DP-RCNet block, we denote its input as $U_b \in R^{D_r \times I \times J}$. The first RNN process each chunk $U_b(:,:,j) \in R^{D_r \times I}, j = 1, \ldots, J$ separately to learn local information. The bi-directional long-short time memory (Bi-LSTM) [28] is adopted to learn from both directions and to better control the information flow. Since the feature dimension is doubled by the Bi-LSTM layer, i.e., the output of first RNN has the shape of $2D_r \times I \times J$, an extra fully connected layer is needed to recover the dimension back to $D_r \times I \times J$. The 2D convolutional layers are applied similarly with their counterparts in the original ConvNeXt block, which means that the first one expands the feature dimension to $4D_r$ and the second one brings it back. The second RNN process its input similarly but on the inter-chunk dimension to model global dependency. Besides, the skip connection is used to enhance the gradient flow.

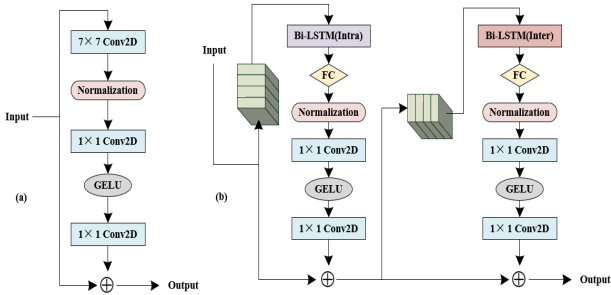

Figure 3: (*a*) *The original ConvNeXt block and* (*b*) *the proposed DP-RCNet block.*

### 2.2.3. Overlap-add module

The output of the backbone module $\widetilde{U}$ has the dimension $D_4 \times I \times J$ and is non-linearized with a GELU function. Since the separator needs to estimate a mask for each speaker, the obtained feature tensor is further fed into a 2-D convolutional layer with kernel size $1 \times 1$ to change the feature dimension from $D_4$ to $SN$. The output stacked tensor $V \in R^{SN \times I \times J}$ is transformed back to $S$ sequences of original shape $N \times K$ through the overlap-add module. These outputs are further passed through the sigmoid function to obtain the estimated masks for all $S$ speakers.

## 3. Experimental setup

### 3.1. Datasets

To evaluate our proposed network, we use three common datasets: WSJ0-2Mix [5], Libri2Mix [29] and WSJ0 Hipster Ambient Mixtures (WHAM!) [30]. WSJ0-2Mix contains 30h training set, 10h validation set and 5h test set. Each mixture is generated by randomly selecting utterances from different speakers in Wall Street Journal dataset (WSJ0), which are further mixed at a random signal-to-noise ratio (SNR) between -5dB and 5dB. Libri2Mix is an open-source dataset for single channel speech separation. train-100, dev and test set in Libri2Mix are used in this paper as training, validation and test set, respectively. WHAM! extends WSJ0-2Mix towards more realistic and challenging scenarios. Two speaker mixtures from WSJ0-2Mix are further mixed with real ambient noise, which are collected in coffee shops, restaurants and bars. All speech signals used are resampled at 8 kHz to reduce computational complexity and memory consumption.

### 3.2. Model configurations

For both encoder and decoder, the frame length $L$ is set to 16 samples and a 50% stride size is used. The feature dimension $N$ is 256 and the reduced feature dimension is 64, i.e.: $B = 64$. For the segmentation module and overlap-add module in the separator, the chunk size $I$ is set to 96, which follows the idea proposed in [21] that $I$ should be approximately equal to the square root of the sequence length. The hop size $H$ is set to half of the chunk size $I$. In the backbone module, the feature dimension $D_r$ is 16, 32, 64, 128 for each stage and the number of the DP-RCNet blocks is 3, 3, 9, 3, respectively. The number of hidden units in each Bi-LSTM is set to 128. LayerScale [31] with initial value $10^{-6}$ is used to offer more diversity in the optimization for better convergence. Stochastic depth [32] is adopted to regularize the proposed network.

### 3.3. Training details

All models are trained on 4s long utterances for 150 epochs. Adam [33] is used as the optimization algorithm and the initial learning rate is set to 0.001. For the first 100 epochs, the learning rate is multiplied by a factor of 0.98 for every two epochs. After that, the factor multiplied decreases to 0.9. Gradient clipping is applied with a maximum $L_2$-norm of 5 and automatic mixed-precision is employed to accelerate the training. Early stop is used if there is no improvement for 10 successive epochs on the validation set. The training objective is to maximize the scale-invariant signal-to-distortion ratio (SI-SDR) [34] with the help of utterance-level permutation invariant training (uPIT) [35].

## 4. Results and discussions

For comparison, all results are evaluated through the well-known metrics SI-SDR improvement (SI-SDRi) and SDR improvement (SDRi) [36]. In addition, memory consumption and computational complexity are measured to assess the computational efficiency of various methods.

### 4.1. Results on WSJ0-2Mix dataset

Table 1 shows the results of several well-known methods and our proposed method on WSJ0-2Mix dataset. Moreover, their memory consumption and computational complexity are listed. For fair comparison, memory consumption is evaluated on the same GPU (NVIDIA RTX 3090) for a 4s long utterance using PyTorch profiler [37]. To the best of our knowledge, this is the first work that adapts ConvNeXts to cope with speech separation and our proposed method DP-RCNet, with a stride size 8, achieves 17.7dB and 18.0dB on SI-SDRi and SDRi, respectively. Our proposed method has a stride similar to that of the A-FRCNN-16. Although A-FRCNN-16 can achieve a 0.6dB increase on both metrics and small memory consumption, its computational complexity is almost doubled compared to DP-RCNet. As for DPRNN, the performance of DP-RCNet is comparable to that of DPRNN (stride size 2) with less memory consumption and computational complexity. As indicated in Table 1, although the separation performance increases with the decrease of stride size for DPRNN, its computational efficiency drops dramatically. Besides, the state-of-the-art methods, such as DPTNet and SepFormer, have achieved an impressive performance with SI-SDRi for more than 20dB. However, DPTNet has almost three times the memory consumption and computational complexity of our proposed method since it used both transformer structure and RNNs. As for SepFormer, which entirely adopts transformer structure, although its memory consumption is relatively small with stride size 8, the computational complexity remains high. In addition, the parameter size of our model is significantly smaller than the SepFormer. Therefore, a trade-off between the separation performance and computational efficiency is well achieved.

### 4.2. Results on different datasets

To further verify the applicability of our proposed method, experiments are conducted on Libri2Mix and WHAM! datasets as well. Since these two datasets are more complicated than the classical WSJ0-2Mix dataset, speech separation is much more challenging and the results are shown in Table 2. Conv-TasNet, whose stride is 8, is chosen as the reference method and its performance on these two datasets is reported in [38] and [39], respectively. Our proposed method has achieved good separation performance on both datasets as expected, proving its effectiveness.

Table 2: *Comparison of separation performance on Libri2Mix and WHAM! datasets.*

| Method | Libri2Mix | | WHAM! | |
|---|---|---|---|---|
| | SI-SDRi (dB) | SDRi (dB) | SI-SDRi (dB) | SDRi (dB) |
| Conv-TasNet [17] | 13.2 | 13.6 | 12.7 | — |
| DP-RCNet | 15.1 | 15.5 | 13.8 | 14.2 |

## 5. Conclusions

In this paper, we proposed a new time-domain speech separation network named DP-RCNet. The network structure was inspired by the newly proposed ConvNeXts and the well-known dual-path strategy. Embedding two RNNs in the ConvNeXt block enables the network to learn efficiently the local information and global dependency. With the four-staged structure, the feature dimension is doubled for every successive stage and the separation is performed gradually at finer and finer scales. The experimental results prove the effectiveness of the proposed method in both clean and noisy scenarios. In addition, a trade-off between the separation performance and the computational efficiency is achieved.

## 6. Acknowledgements

The work was supported by the National Natural Science Foundation of China (Grant No. 61831019).

Table 1: *Comparison of separation performance on WSJ0-2Mix dataset and evaluation of computational efficiency*

| Method | #Params (M) | Stride (samples) | WSJ0-2Mix SI-SDRi (dB) | SDRi (dB) | Memory Consumption (GB) | MACs (G) |
|---|---|---|---|---|---|---|
| DPCL++ [5] | 13.6 | 64 | 10.8 | — | — | — |
| uPIT-blstm-ST [35] | 92.7 | 128 | — | 10.0 | — | — |
| ADANet [7] | 9.1 | 64 | 10.4 | 10.8 | — | — |
| BLSTM-TasNet [14] | 23.6 | 20 | 13.2 | 13.6 | — | — |
| Conv-TasNet [17] | 5.1 | 8 | 15.3 | 15.6 | 0.78 | 20.8 |
| SuDoRM-RF 1.0x [19] | 2.7 | 10 | 17.0 | — | — | — |
| FurcaNeXt [18] | 51.4 | — | — | 18.4 | — | — |
| A-FRCNN-16 [20] | 6.1 | 10 | 18.3 | 18.6 | 0.76 | 140.3 |
| DPRNN (With various strides) [21] | 2.6 | 8 | 16.0 | 16.2 | 1.36 | 21.58 |
| | | 4 | 17.0 | 17.3 | 2.66 | 42.53 |
| | | 2 | 17.9 | 18.1 | 5.29 | 85.01 |
| | | 1 | 18.8 | 19.0 | 10.51 | 169.21 |
| DPTNet [22] | 2.7 | 1 | 20.2 | 20.6 | 14.74 | 209 |
| SepFormer [23] | 26 | 8 | 20.4 | 20.5 | 2.01 | 126.23 |
| DP-RCNet | 9.2 | 8 | 17.7 | 18.0 | 4.23 | 76.63 |